\documentclass{article}
\usepackage{arxiv}

\usepackage{multirow}
\usepackage{tabu}
\usepackage{url}
\usepackage{verbatim} 

\usepackage[utf8]{inputenc} 
\usepackage[T1]{fontenc}    
\usepackage{hyperref}       
\usepackage{url}            
\usepackage{booktabs}       
\usepackage{amsfonts}       
\usepackage{nicefrac}       
\usepackage{microtype}      
\usepackage{lipsum}		
\usepackage{graphicx}

\usepackage{lineno}
\graphicspath{ {./figs/} }

\begin{document}


\title{Dense Retrieval for Low Resource languages - the Case of Amharic Language}

\author{
 Tilahun Yeshambel \\
 Addis Ababa University \\
 Addis Ababa, Ethiopia \\
 \texttt{tilahun.yeshambel@uog.edu.et} \\
 \And
 Moncef Garouani \\
 Univ. Capitole, IRIT, UMR5505 CNRS \\
 Toulouse, France \\
 \texttt{moncef.garouani@irit.fr} \\
 \And
 Serge Molina \\
 IRIT, UMR5505 CNRS, Univ. de Toulouse \\
 Toulouse, France \\
 \texttt{serge.molina@irit.fr} \\
 \And
 Josiane Mothe \\
 INSPE, UT2J, IRIT, UMR5505 CNRS \\
 Toulouse, France \\
 \texttt{josiane.mothe@irit.fr} \\
}





\maketitle


\begin{abstract}
This paper reports some difficulties and some results when using dense retrievers on Amharic, one of the low-resource languages spoken by 120 millions populations. The efforts put and difficulties faced by Univ. Addis Ababa Univ. toward Amharic Information Retrieval will be developed during the presentation.
\end{abstract}

\section{Context and Objectives}
\label{Sec:Introduction}
Amharic, the official language of Ethiopia and the second most spoken Semitic language after Arabic, presents significant challenges for information retrieval due to its complex morphology, vocabulary mismatches, and limited linguistic and computational resources. The language's rich inflectional and derivational structures result in high lexical variability, making term-based matching difficult. Additionally, the scarcity of large annotated datasets, pre-trained retrieval models, and essential NLP tools further complicates the development of effective retrieval systems.

Traditional retrieval models like BM25 ~\cite{robertson1995okapi}, which rely on exact term matching, struggle to bridge the semantic gap caused by Amharic’s morphological complexity. Shallow neural embeddings such as word2vec and FastText have been explored as alternatives, showing promise in capturing semantic relationships. However, their effectiveness remains limited due to the lack of high-quality, large-scale training corpora. More advanced retrieval techniques, including sparse expansion models like SPLADE and late interaction models like ColBERT~\cite{santhanam2021colbertv2}, have demonstrated strong performance in high-resource languages but remain largely unexplored for Amharic. These models require substantial data, 
which are not readily available for low-resource languages.

This work investigates the feasibility of ColBERT for Amharic retrieval by evaluating their performance under low-resource constraints. We develop an Amharic-specific dataset to facilitate training, and assess the trade-off between traditional sparse methods and more dense retrieval models. By addressing these challenges, we aim to advance the capabilities of retrieval systems for low-resource languages and contribute to more inclusive and effective information access.

\section{Datasets and Rankers}
\label{Sec:Evaluation}
We used three datasets:

\textbf{2AIRTC} ~\cite{yeshambel20202airtc} is an Amharic ad-hoc information retrieval test collection consisting of more than 10,000 documents  covering various domains such as news, government publications, and academic content, and  
originate from multiple sources, including Amharic Wikipedia, news websites, blogs, and social media platforms. The dataset contains 240 manually curated queries and human-annotated relevance judgments.

\textbf{Train Amharic} is a data set developed by the same authors for training dense retrievers like  ColBERT, consisting of 152 queries, 152 documents, and corresponding query relevance annotations. 

\textbf{AfriCLIRMatrix} is a  cross-lingual IR dataset that includes queries in English and documents in 15 African languages. For Amharic, it contains 15'458 documents, 248'672 queries and 264'690 judgments (a document is relevant if retrieved by BM25 according to the authors)  along with 1,500 test queries and 1,582 test judgments~\cite{ogundepoAfriCLIRMatrixEnablingCrossLingual2022}. We translated the test queries from English to Amharic using the NLLB-200 model ~\cite{nllbteamNoLanguageLeft2022a}.

On rankers, we used:

\textbf{BM25}  a well-established probabilistic ranking function. 
It is is known for its efficiency and robustness in keyword-based retrieval. 
We used the Lemur project Indri ~\cite{strohman2005indri} 5.21 implementation with default parameters and   Amharic stopword list ~\cite{yeshambel2020Stop}.

 \textbf{ColBERT} (Contextualized Late Interaction for Efficient Document Retrieval)~\cite{santhanam2021colbertv2} is a late interaction neural retrieval model that leverages BERT-based token embeddings for finer-grained semantic matching. 
ColBERT computes contextualized token embeddings for both query and document, allowing per-token similarity computation at retrieval time. 
We employ the ColBERTv2 variant.~\footnote{\url{https://github.com/stanford-futuredata/ColBERT }}

ColBERT is BERT-based. 
We used different masked language models pre-trained on the same Amharic datasets consisting of Oscar Amharic \cite{ortiz-suarez-etal-2020-monolingual,OrtizSuarezSagotRomary2019}, C4 ~\cite{raffel2020exploring}, and the Amharic Sentences Corpus ~\cite{yimam2021introducing}, totalling at 290M tokens. We use models produced independently from our research \url{https://huggingface.co/rasyosef/[ModelName]}: bert-medium-amharic which creates token level 512 dimension representation,  roberta-base-amharic, a sentence transformer based on the XML Roberta architecture ~\cite{conneau2019unsupervised} creating 768 dimension representations at a sentence level, and roberta-amharic-text-embedding-base, finetuned from roberta-base-amharic.

\section{Results and Discussion}
\begin{figure}
    \centering
    \includegraphics[width=0.45\linewidth]{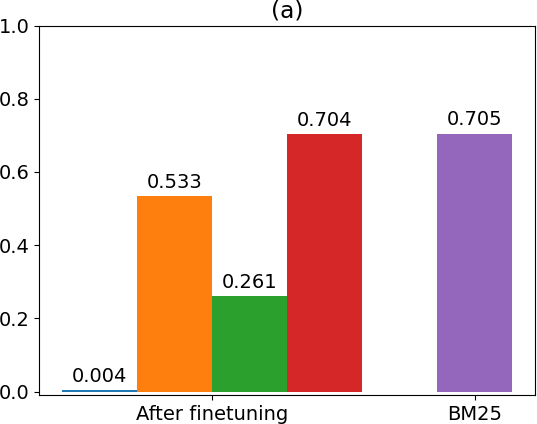}
    \includegraphics[width=0.45\linewidth]{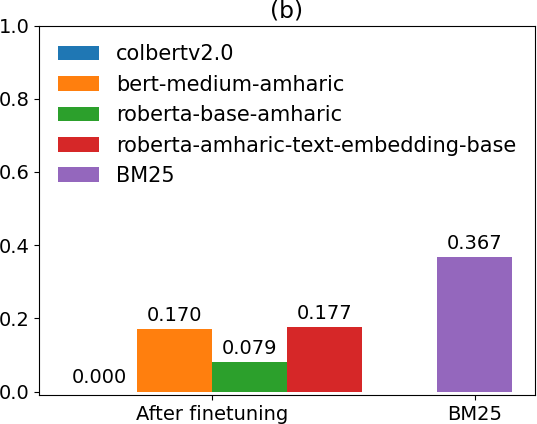}
    \caption{\textbf{NDCG@10}  (a) 2AIRTC (b) AfriCLIRMatrix - First bars are without specific Amharic pre-training latest with.}
    \label{fig:NDCG@10}
\end{figure}

When using ColBERT with English-based BERT models, the results are very poor (as expected) with NDCG@10 values from 0.000 to 0.061 on both collections. 
When using BERT models pre-trained with Amharic, before fine tuning on the Training Amharic collection, the results remain very poor, from 0.000 to 0.005 across both test collections on ColBERT.

In contrast, Figure~\ref{fig:NDCG@10} presents the results when using BERT models after fine tuning on the Train Amharic collection. For 2AIRTC, the results are generally better than for AfriCLIRMatrix with ColBERT best performing model (roberta-
amharic-text-embedding-base) achieving an NDCG@10 value of  0.704 for 2AIRTC and 0.177 for AfriCLIRMatrix, the models are following the same performance ranks on Precision@10, Recall@10 and MRR@10 metrics.

The best performing model using ColBERT is a sentence transformer RoBERTa model, which produces a single sentence-level representation and does not benefit from ColBERT late interaction system. When comparing to the bert-medium-amharic model -second best performing model- and the roberta-base-amharic model, 
the results indicate the performance advantage is not coming from the architecture but rather the additional finetuning that has been done. We also tried SPLADE (Sparse Lexical and Expansion Model for Document Retrieval)~\cite{lassance2024splade} which applies transformer-based token expansion to improve document-query matching. It expands input tokens into weighted sparse representations over a vocabulary space,  guided by a BERT-based encoder.
We used the DistilSPLADE v2 variant and tried the same BERT and RoBERTa  variants as for ColBERT but could not achieve non-null results.

Computation times (see Table ~\ref{tab:execution_times}) are many orders of magnitude larger when indexing with ColBERTv2 best performing model compared to Lemur Indri, amounting to 29 times (2AIRTC) and 46 times (AfriCLIRMatrix) Indri's indexing time. 
Retrieval times using ColBERTv2 are close to BM25 with Indri, amounting to 0.78 times (2AIRTC) and 1.17 times (AfriCLIRMatrix). 
Dense retrieval models 
can be competitive with BM25. 
We show that 150 training examples are enough, 
which is lower than previously reported 
(1000) ~\cite{karpukhin2020dense} and can be crafted manually, allowing broader access to IR in low-resource contexts.
The main bottle neck is MLM pretraining -are pre-training one to evaluate this step. With low resource languages, it is likely that the MLM has been trained on the same documents than the collection used to evaluate IR, maybe almost exclusively; which can be seen as a kind of collection contamination~\cite{kalal2024training}.

In many low-resource language settings, the availability of computational infrastructure can be a major limiting factor in developing advanced retrieval models. Ethiopia, despite its growing digital landscape, faces significant challenges in accessing high-performance computing resources necessary for training and fine-tuning deep learning models. 
This work was made possible through access to French computational resources. 
The reliance on external resources highlights the urgent need for investment in local AI infrastructure to support research in low-resource languages.
\begin{table}[h]
    \centering
    \begin{tabular}{lc|cc|cc}
         & & \multicolumn{2}{c|}{2AIRTC} & \multicolumn{2}{c}{AfriCLIRMatrix} \\
        Model & Train & Index & Retrieve & Index & Retrieve \\
        \hline
        colbertv2.0 & 00:36 & 06:24 & 00:25 & 04:44 & 00:51 \\
        bert-amh & 00:29 & 04:30 & 00:34 & 02:51 & 01:10 \\
        rbert-amh & 00:38 & 06:38 & 00:20 & 05:25 & 00:42 \\
        rbert-amh-emb & 00:35 & 06:24 & 00:31 & 05:24 & 01:06 \\
        \hline
        BM25 $\square$ & --- & 00:13 & 00:40 & 00:07 & 00:56 \\
        \hline
    \end{tabular}
    \caption{Execution times for each retrieval models (mm:ss) first rows are for ColBERTv2 with the various BERT/ColBERT models
    ; last row is Lemur Indri; }
    \label{tab:execution_times}
\end{table}



\bibliographystyle{unsrt}  
\bibliography{arxiv_SIGIR_Low_Resource_Soumis_V1}

\end{document}